\documentclass[aps,onecolumn,notitlepage,superscriptaddress]{revtex4-1}

\usepackage{physics}
\usepackage{graphicx}

\begin{document}
\title{Superconducting optical response of photodoped Mott insulators}
\author{Jiajun Li}
\email{cong.li@fau.de}
\affiliation{Department of Physics, University of Erlangen-Nuremberg, 91058 Erlangen, Germany}
\author{Denis Golez}
\affiliation{Center for Computational Quantum Physics, Flatiron Institute, 162 Fifth Avenue, New York, NY 10010, USA}
\author{Philipp Werner}
\affiliation{Department of Physics, University of Fribourg, 1700 Fribourg, Switzerland}
\author{Martin Eckstein}
\affiliation{Department of Physics, University of Erlangen-Nuremberg, 91058 Erlangen, Germany}

\begin{abstract}Ultrafast laser pulses can redistribute charges in Mott insulators on extremely short time scales, leading to the fast generation of photocarriers. It has recently been demonstrated that these photocarriers can form a novel $\eta$--paired condensate at low temperatures, featuring a staggered superconducting pairing field. In this conference paper, we discuss the origin of the $\eta$--paired hidden phase and its optical response which may be detected in a pump-probe experiment. The hidden phase may be relevant for possible light-induced superconductivity in Mott insulators.
\end{abstract}

\maketitle

\section{introduction}

The advancement of ultrafast laser techniques allows to access highly excited states in solid state systems and holds the promise of revealing novel states of matter in non-equilibrium regimes. In this field, tantalizing findings, including the ultrafast manipulation of long-range orders and the observations of hidden phases in complex transition metal oxides and chalcogenides, have been the subject of intense research \cite{ichikawa2011,stojchevska2014,li2018nat,golez2017}. Recently, the possibility of light-induced superconductivity has attracted considerable interest from both experimental and theoretical groups \cite{fausti2011,mitrano2016,denny2015,raines2015,patel2016,okamoto2016,kennes2017,sentef2017,babadi2017}. It has been shown that, under a strong laser pulse, the generation of photocarriers in a Mott insulator can be accompanied by a strong enhancement of its reflectivity, suggesting the apparent formation of a non-thermal superconducting phase. Multiple theoretical proposals have suggested  a variety of scenarios in which the superconductivity can be enhanced through external driving. However, these scenarios often apply to relatively special models, and it is still under debate whether they actually explain the origin of the superconducting-like behaviors observed in experiments. 

Among the various scenarios, an interesting possibility is that the observed phenomenology is related to 
a certain class of 
excited states of the fermionic Hubbard model, the $\eta$--paired states, which are eigenstates of the $\eta$--pseudospin \cite{yang1989}. Intuitively,  the $\eta$--pseudospin describes the charge degrees of freedom of the repulsive Hubbard model, 
in analogy to
the spin angular momentum $\mathbf{S}$ describing the spin degrees of freedom. External driving can excite the charge sector and populate these $\eta$--paired states, enhancing the phase stiffness of the system. This has already been confirmed by exact diagonalization in small systems \cite{kaneko2019}. In large systems, these coherent excited states can be vulnerable to dissipation and thermalization, unless they are protected by symmetry under special circumstances \cite{tindall2019}. 

In this conference paper,  we will discuss the general optical properties of a photodoped single-band Mott insulator. We will first discuss the formation of a hidden $\eta$--paired phase. This is due to a general pairing mechanism between the photoinduced charge carriers. We then demonstrate that the $\eta$--paired phase consistently exhibits a superconducting optical response.

\section{The formation of cold photodoped states}

\begin{figure}
\includegraphics[scale=0.8]{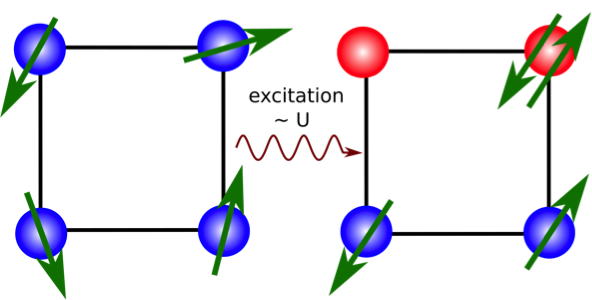}
\caption{Creation of photocarriers through photoexcitation.}
\label{excitation}
\end{figure}

 Under resonant photoexcitation, a Mott insulator can be turned into a metal through the generation of charge excitations. We exemplarily consider a single-band fermionic Hubbard model,
 \begin{align}
 H=-t\sum_{\langle ij \rangle\sigma}e^{i\phi_{ij}}d^\dag_{i\sigma} d_{j\sigma}+U\sum_i n_{i\uparrow}n_{i\downarrow},
 \label{ham}
 \end{align}
 where $\phi_{ij}(t)=\int_{\mathbf{R}_i}^{\mathbf{R}_j} d\mathbf{r}\cdot \mathbf{A}(\mathbf{r},t)$ denotes the Peierls phase. For vector potential $\mathbf{A}=0$ and half-filling $n=1$, the equilibrium ground state is 
 insulating and exhibits
 two Hubbard bands in the local density of states. The chemical potential lies inside the Mott gap, which is proportional to the interaction parameter $U$. A light pulse corresponds to an oscillating 
 vector potential $\mathbf{A}(t)$,  which can be treated as uniform in space for optical frequencies, where the associated magnetic field is negligible. 
 A resonant pulse couples the two Hubbard bands and transfers electrons from the lower band to the upper band. In real space, this corresponds to incoherent creation of doubly occupied sites $\ket{\uparrow\downarrow}$ (doublons) and empty sites $\ket{0}$ (holons). This process is termed photodoping and the charge excitations act as charge carriers in the resulting metallic photodoped state. Fig.~\ref{excitation} is a schematic illustration of the process.
 
In experiments, the creation of photodoped states can be very fast (just a few femto-seconds), corresponding to the time scales of the electric pulse and the electronic dynamics in the system. This state generally decays back to a thermal state through recombination of the charge excitations, 
but the time scale of this recombination can easily be thousands of hopping times, as confirmed both in solids and in cold atom systems \cite{iwai2003,sensarma2010,iyoda2014}. This can be pico-seconds in solids and even much longer in cold atom systems.
In other words, since the recombination of a doublon and a holon is associated with the dissipation of a large amount of energy $U$, it requires multiple scattering events, leading to a lifetime which scales exponentially with $U$. This hierarchy of time-scales leads to a prethermal time regime, in which the charge excitations are present but have partially thermalized within the upper and lower Hubbard bands 
\cite{werner2012,eckstein2013,mor2017,ligges2018,peronaci2019}. This metastable state appears to be distinct from any thermal states, due to the presence of both excess doublons and holons. Furthermore, the large energy injected by the external driving can be `hidden' in the potential energy $U$ of doublons and holons, 
and the kinetic energy of these charge excitations can be small and can be further reduced due to dissipation into environments \cite{eckstein2013, peronaci2019} or by an evaporative cooling protocol \cite{werner2019nat}.
This can in principle lead to a low effective temperature of the photodoped state.

\begin{figure}
\includegraphics[scale=1]{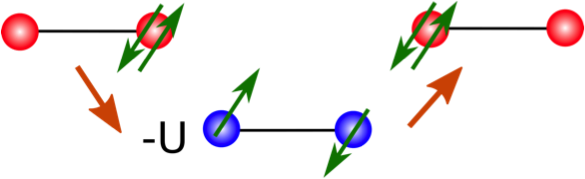}
\caption{The doublon-holon exchange interaction.}
\label{exchange}
\end{figure}

It is worth noting that, due to strong intra-band thermalization effects, this photodoped state should show universal physical properties independent of the non-equilibrium protocol used to create the state \cite{werner2012}. To describe this state, we can start with Eq.~\eqref{ham} and project out the charge recombination processes \cite{li2019,bukov2016,eckstein2017}. This results in an effective hamiltonian with two mixed liquids,
\begin{align}
H_{\rm eff}=-J_{\rm ex}\sum_{\langle ij\rangle}\boldsymbol{\eta}_i\cdot\boldsymbol{\eta}_j+J_{\rm ex}\sum_{\langle ij\rangle}\mathbf{S}_i\cdot\mathbf{S}_j-t\sum_{\langle ij\rangle\sigma}\Big[\mathcal{P}_i d^\dag_{i\sigma} d_{j\sigma}\mathcal{P}_j+\bar{\mathcal{P}}_i d^\dag_{i\sigma} d_{j\sigma}\bar{\mathcal{P}_j}\Big],
\label{heff}
\end{align}
where $\mathcal{P}=n_\uparrow n_\downarrow+(1-n_\uparrow) (1-n_\downarrow)$ projects to the doublon and holon subspace spanned by $\{\ket{\uparrow,\downarrow},\ket{0}\}$, and $1-\mathcal{P}=\bar{\mathcal{P}}$. In the first term, we have defined the $\eta$ pseudospins as follows,
\begin{align}
\eta^+_i&=\eta^x_i+i\eta^y_i=(-1)^i d^\dag_{i\uparrow} d^\dag_{i\downarrow}=\eta^{-\dag}_i\nonumber\\
\eta^z_i&=\frac{1}{2}(n_i-1).
\end{align}

$\eta^\pm$ is essentially a staggered superconducting pairing field, while $\eta^z$ measures the local charge density fluctuation. This term describes an attraction between doublons and holons and only locally acts on the doublon-holon subspace. It can be associated with a virtual process involving the creation of singlon states, see Fig.~\ref{exchange}. On the other hand, the local spin angular momentum operator $S^\mu_i=\frac{1}{2}d^\dag_{i\alpha}\sigma^\mu_{\alpha\beta}d_{j\beta}$ only acts on the singlon subspace $\{\ket{\uparrow},\ket{\downarrow}\}$. The second term is simply the superexchange interaction between two neighboring sites. The third term describes the exchange of neighboring singlon and doublon-holon states. This model is a generalization of the $t$--$J$ model in equilibrium and its general properties can be difficult to extract. One may exclude a coherent superposition of singlon and doublon-holon states at the same site, and in the large coupling $U\gg t$ limit, the system should exhibit the demixing of doublon-holon and singlon liquids, which are dominated by doublon-holon pairing ($\boldsymbol{\eta}_i\cdot\boldsymbol{\eta}_j$) and antiferromagnetic correlations ($\mathbf{S}_i\cdot\mathbf{S}_j$), respectively. For a large population of doublon-holon pairs, the system should thus exhibit enhanced $\eta$--pairing and even an $\eta$--paired phase with non-zero order parameter $\langle\eta^{x,y}\rangle$, i.e., a non-zero staggered superconducting pairing field.

\section{The $\eta$--paired hidden phase}

To study the quantitative properties of the photodoped state, we use non-equilibrium dynamical mean-field theory (DMFT) to examine the excited Hubbard model on a Bethe lattice with infinite coordination number. A strong-coupling expansion limited to the lowest order (non-crossing approximation) is used to solve the associated Anderson impurity model. With phonon coupling, the system driven by a chirped electric pulse shows a significantly enhanced $\eta$--pairing susceptibility \cite{li2019}. Furthermore, with an evaporative cooling protocol, one can reach a cold photodoped state, allowing for a fast formation of the $\eta$--paired phase \cite{werner2019prb}. 

\begin{figure}
\includegraphics[scale=1.2]{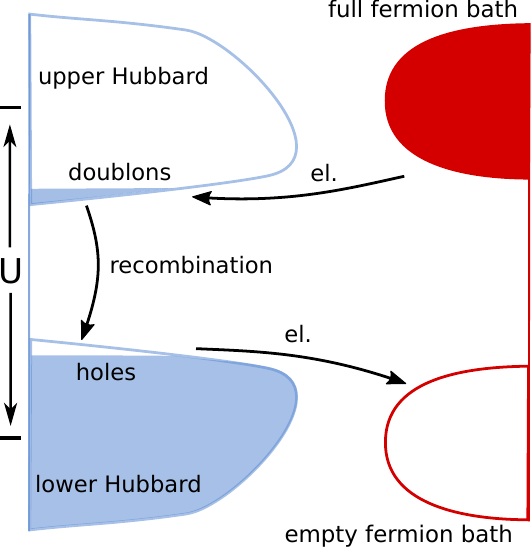}
\caption{The bath-doping protocol.}
\label{protocol}
\end{figure}

A more systematic study can be carried out through a bath-coupling protocol. By coupling the Hubbard lattice to two separate fermion baths, one can simultaneously inject doublons and holons into the ground state of the system. The protocol is illustrated in Fig.~\ref{protocol}. As long as the bath coupling is small, the physical properties of the state should be minimally affected. On the other hand, due to the small recombination rate of charge excitations, a weak bath coupling suffices to produce a large amount of doublons and holons in the system.

\begin{figure}
\includegraphics{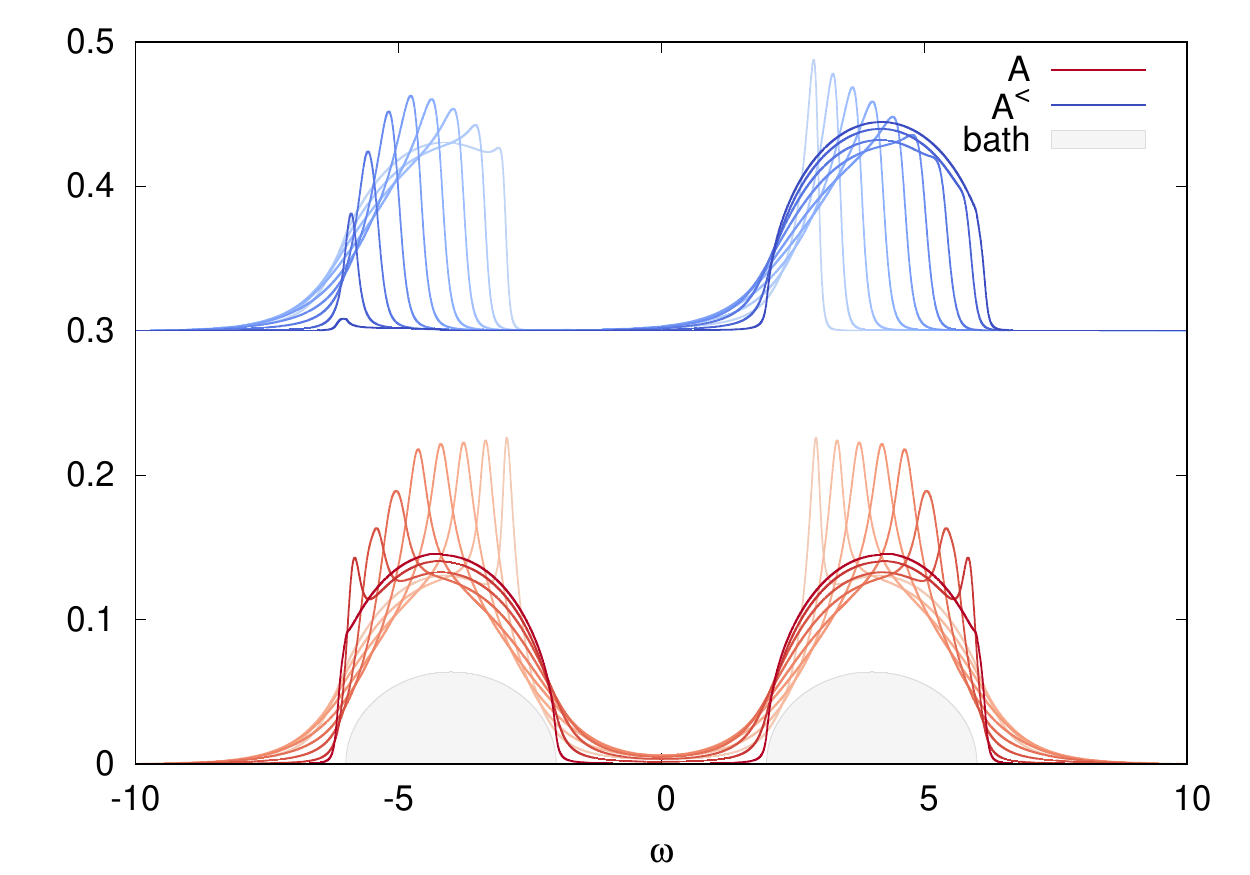}
\caption{Photodoped states created through bath-coupling. The grey shaded curve is the density of states of the fermion baths. By changing chemical potential, one obtains different density of states (red curve, from light to dark) and occupation (blue curve, from light to dark). The chemical potential is from $\mu_U=3.0$ to $\mu_U=6.0$ with stepsize $0.2$. Note that $\mu_U=-\mu_L$.}
\label{specs}
\end{figure}

Specifically, we consider a bath coupling of the form
\begin{align}
H_{\rm coupl}=g\sum_{i\alpha\sigma} (c^\dag_{i\alpha\sigma} d_{i\sigma}+{\rm h.c.})+\sum_{i\alpha\sigma}\epsilon_{\alpha}c^\dag_{i\alpha\sigma}c_{i\alpha\sigma}, 
\end{align}
with semi-elliptic densities of states for the baths $D(\omega)\propto\sqrt{1-(\omega-V_s)^2/W^2}$ of bandwidth $2W=4t$. Two fermionic baths $s=U,L$ are considered with $V_U=U/2$ and $V_L=-U/2$. The chemical potential $\mu_U=-\mu_L$ and the bath temperature $T_b$ can be changed to implicitly control the density of doublon-holon pairs (measured by double occupancy $d=n_\uparrow n_\downarrow$) and effective temperature of doublon-holon liquid $T_{\rm eff}$.  The spectral function and occupation is shown for increasing chemical potential $\mu_U$ in Fig.~\ref{specs}. In practice,  $T_{\rm eff}$ is measured by fitting the distribution function $f(\omega)=-\operatorname{Im} G^<(\omega)/2\operatorname{Im} G^r(\omega)$ at the two effective Fermi surfaces of the doublon and holon separately.

With $\Gamma=g^2/W^2=0.1$, one can observe the emergence of a nonzero order parameter $\langle \eta^x\rangle$ at about $d\sim 0.3$, indicating the formation of an $\eta$--paired phase. Moreover, this state is robust to a small next-to-nearest-neighbor hopping, which corresponds to a frustration term for the $\eta$--order. 

In general, various perturbations can reduce the observed $\eta$--paired phase. For example, a next-to-nearest neighbor repulsion $V\sum_{\langle ij\rangle} n_i n_j$ may be translated to a term $4V\sum_{\langle ij\rangle} \eta^z_i\eta^z_j$ in the effective model since $\eta^z\sim\frac{1}{2}(n-1)$, modifying the $\eta$ exchange term to
\begin{align}
-J_{\rm ex}(\eta^x_i\eta^x_j+\eta^y_i\eta^y_j)-(J_{\rm ex}-4V)\eta^z_i\eta^z_j.
\end{align}
When $J_{\rm ex} < 4V$, this may reduce the $\eta$--pairing in the XY-plane while inducing a charge-density-wave instability.  It is also possible to reach a super-solid phase with the coexistence of the two orders. On the other hand, this repulsion reduces the trend of the system to form segregated doublon-rich and holon-rich regions, which should be beneficial to the $\eta$--paired phase. 

\begin{figure}
\includegraphics{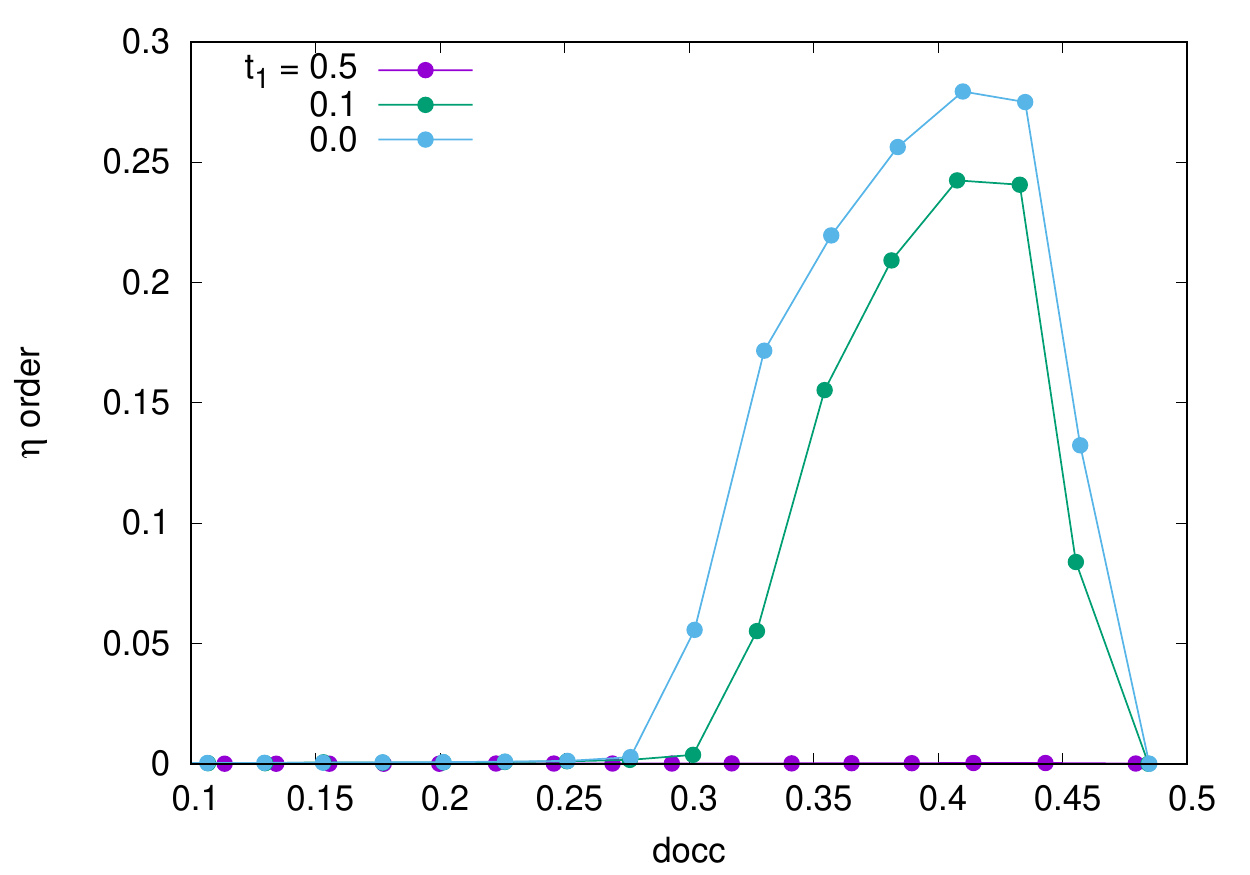}
\caption{Evolution of $\eta$--order under rising double occupancy. A small next-to-nearest neighbor hopping $t_1$ reduces the $\eta$--order.}
\label{eta}
\end{figure}

On the other hand, the electron-phonon interaction may help cooling the doublon-holon liquid, which favors the paired phase, but can also give rise to an effective retarded attraction between electrons, which may favor the spatial separation of doublons and holons. The final fate of the $\eta$--paired phase 
is therefore expected to depend on the details of the model.

\section{Optical response of the $\eta$--paired phase}

The $\eta$--paired phase features non-zero staggered pairing order $\langle\eta^x+i\eta^y\rangle=(-1)^i\langle c^\dag_\uparrow c^\dag_\downarrow\rangle$. However, to confirm that it is truly a superconducting state, it remains to be clarified how it couples to electromagnetic fields as well as  the resulting optical response that is experimentally detectable. Indeed, the $\eta$--condensate is essentially a Bose-Einstein condensate of the doublon-holon liquid. It has been argued that, due to the fact that its excitation spectrum is of ferromagnetic nature \cite{rosch2008}, the 2D $\eta$--condensate should \emph{not} exhibit superfluidity. 

\begin{figure}
\includegraphics[scale=0.65]{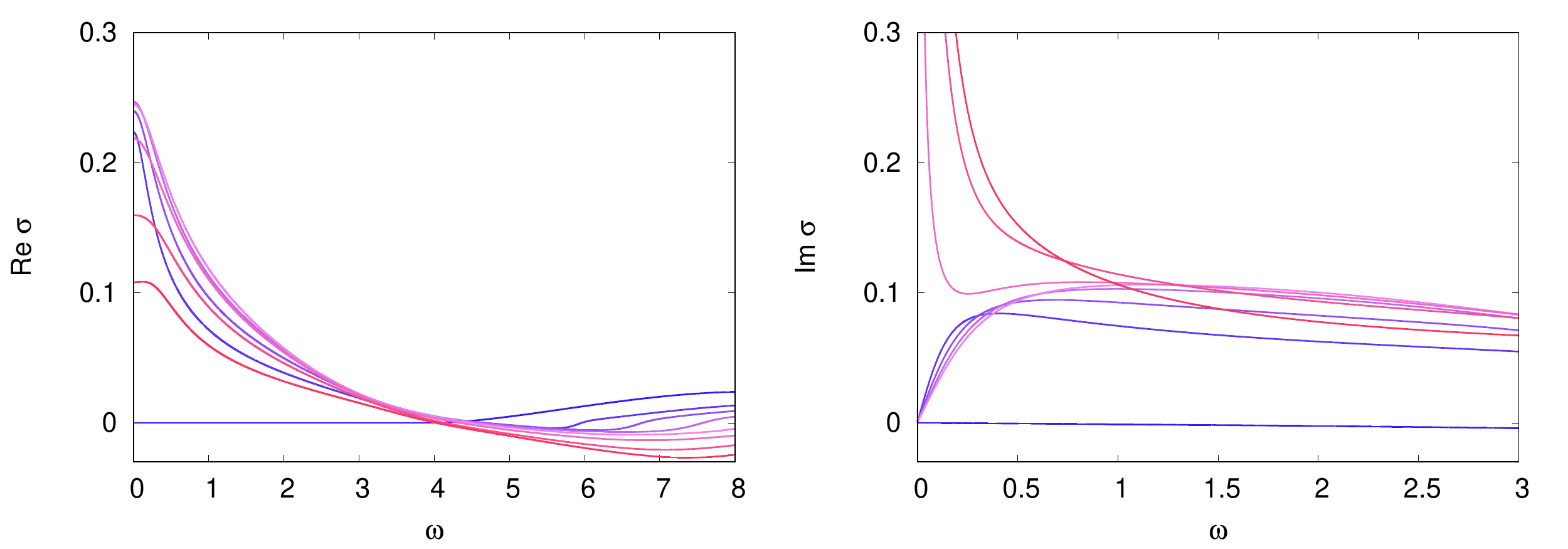}
\caption{Optical conductivity obtained from DMFT. The chemical potential rises from $\mu_U=3.0$ to $\mu_U=5.4$ with stepsize $0.4$ (from blue to red). The resulting double occupancy is from $d\approx0.11$ to $d\approx0.41$.
}
\label{oc}
\end{figure}

To see the coupling of the $\eta$--condensate to electromagnetic fields, it is instructive to note that, under the Peierls substitution,
\begin{align}
d_{i\sigma}^\dag&\to e^{i\int^{\mathbf{R}_i} d\mathbf{r}\cdot\mathbf{A}} d_{i\sigma}^\dag,\nonumber\\
\eta^\pm_i&\to e^{\pm2i\int^{\mathbf{R}_i} d\mathbf{r}\cdot\mathbf{A}}\eta^\pm_i,
\end{align}
which gives the minimal
extension of the Hamiltonian Eq.~\eqref{heff} with a gauge-invariant coupling to classical electromagnetic fields. 
This should be qualitatively correct at least for slowly varying and weak electric field.
%
Under fast and strong electromagnetic driving, the model parameters such as $J_{\rm ex}$ should also be renormalized \cite{aoki2014}. In the special case where $\mathbf{A}$ is constant both in time and space, the above substitution is formally equivalent to the Peierls substitution in \eqref{ham}. This case is relevant to the superconducting optical response since a constant $\mathbf{A}$ can be viewed as an electric pulse satisfying $\mathbf{A}=-\int dt \mathbf{E}(t)$ at infinite past. Hence, the $\eta$ exchange term becomes 

\begin{align}
-2J_{\rm ex}\sum_{\langle ij\rangle}e^{2i\phi_{ij}}\eta^+_i\eta^-_j+\ldots,
\end{align}
where $\phi_{ij}$ is the Peierls phase. In the symmetry-broken phase with nonzero $\eta$--order, this term contributes a current component $\mathbf{j}=\delta H / \delta A \approx -4J_{\rm ex}\eta^2 \mathbf{A}$, in which a mean-field decoupling is assumed. This is the superconducting current associated with the $\eta$--condensate. The superconducting Drude weight 
is 
$D=4J_{\rm ex}\eta^2\propto \eta^2/U$.

In the following we show the optical conductivity $\sigma(\omega)=\chi_{JJ}(\omega)/(\omega+i0^+)$ in the DMFT solution, where $\chi_{JJ}\equiv\delta j/\delta A$ is the current correlation function \cite{eckstein2008}. As the double occupancy $d$ increases, a broad Drude peak gradually grows out of the equilibrium Mott insulating gap. At large photodoping, a significant negative optical conductivity develops for frequency $\omega\sim U$ in Fig.~\ref{oc}. This negative optical conductivity can be explained by the population of doublons and holons: under a resonant probe pulse, charge recombination (doublon-holon recombination) may occur and release energy back to the driving force, resulting in negative ``Joule heating". It can be seen that the calculated optical conductivity satisfies the $f$-sum rule $\int d\omega\sigma(\omega)=-\pi E_K/4$. 

\begin{figure}
\includegraphics[scale=0.65]{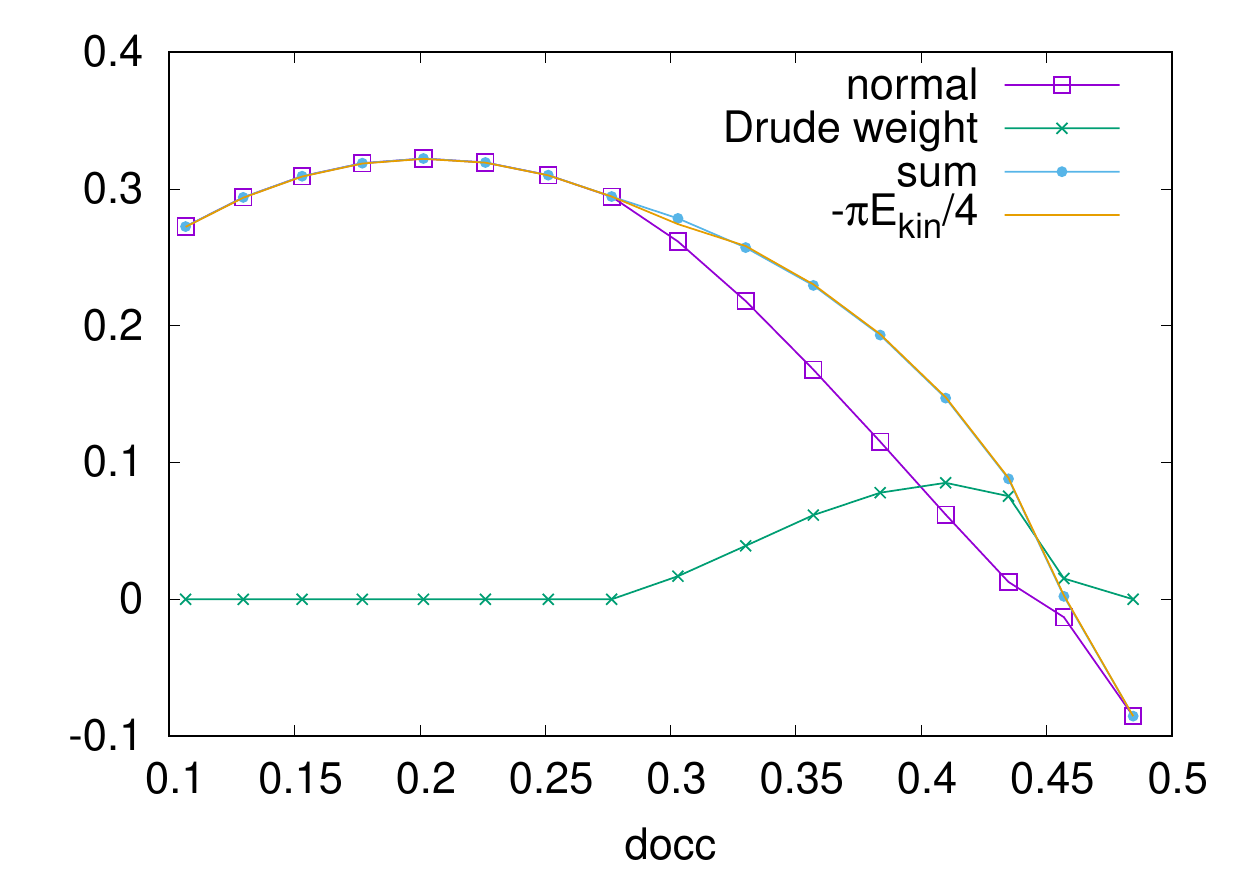}
\caption{The emergence of a SC Drude weight $D=-\chi_{JJ}(0)$ and test of the $f$-sum rule.}
\label{drude}
\end{figure}

When the system enters the $\eta$--paired phase, an ideal Drude peak, i.e., a delta function peak $\pi D\delta(\omega)$, appears in $\operatorname{Re}\sigma$, see Fig.~\ref{drude}. This confirms the zero-resistivity effect. On the other hand, the $1/\omega$ divergence appears in the imaginary part of the conductivity, imposed by analyticity. Furthermore, this behavior corresponds to $\chi_{JJ}(0)\ne0$, which implies London's equation $j=-D A$ consistent with the analytical result. Thus, we have also confirmed the Meissner effect in the $\eta$--paired phase. The scaling $D\propto \langle\eta\rangle^2/U$ has also been numerically confirmed \cite{li2019}.

\section{conclusion}
In this conference paper, we have discussed a ubiquitous phenomenon in the photoinduced dynamics of Mott insulators: photodoping. An ultrafast laser pulse can quickly create charge excitations, which take orders of magnitudes longer time to decay. The relative long-livedness of charge excitations permits an approximate treatment of the excited Mott insulator in the prethermal time regime, where the intra-band thermalization of the excess doublons and holons has taken place, while the charge recombination and full thermalization still require much longer times. In this regime, it is argued that a novel superconducting phase can form due to the intrinsic doublon-holon pairing mechanism. This effect may be enhanced or suppressed by other perturbations and interactions in a realistic system. The interplay between the $\eta$--pairing mechanism and multiple local orbitals as well as nontrivial band topology can be a promising subject reserved for future studies.

\begin{acknowledgments}
We acknowledge discussions with T. Kaneko, O. Parcollet, and A. Millis. M.E. and J. Li were supported by the ERC starting grant No. 716648. PW acknowledges support from ERC Consolidator Grant No. 724103. The Flatiron institute is a division of the Simons foundation.
\end{acknowledgments}
\bibliography{eta.bib}
\end{document}